\documentclass[conference]{IEEEtran}
\IEEEoverridecommandlockouts
\usepackage{cite}
\usepackage{amsmath,amssymb,amsfonts}
\usepackage{algorithmic}
\usepackage{graphicx}
\usepackage{textcomp}
\usepackage{xcolor}
\usepackage{makecell}
\usepackage{tabularx,booktabs,threeparttable}
\def\BibTeX{{\rm B\kern-.05em{\sc i\kern-.025em b}\kern-.08em
    T\kern-.1667em\lower.7ex\hbox{E}\kern-.125emX}}

\usepackage{multirow}

\begin{document}

\title{Emotion-Aware Smart Home Automation\\Based on the eBICA Model\\
\thanks{This work was supported by JSPS KAKENHI Grant Number JP24K20762.}%
\thanks{This paper has been accepted for presentation at the 44th IEEE International Conference on Consumer Electronics (ICCE 2026).}
}

\author{
    \IEEEauthorblockN{
        Masaaki Yamauchi\IEEEauthorrefmark{1}, %
        Yiyuan Liang\IEEEauthorrefmark{1}, %
        Hiroko Hara\IEEEauthorrefmark{1}, %
        Hideyuki Shimonishi\IEEEauthorrefmark{2}, and %
        Masayuki Murata\IEEEauthorrefmark{1}%
    }
    \IEEEauthorblockA{
        \IEEEauthorrefmark{1}Graduate School of Information Science and Technology, %
        The University of Osaka, Osaka, Japan\\
        Emails: \{m-yamauchi, y-liang, murata\}@ist.osaka-u.ac.jp, hara.hiroko.ist@osaka-u.ac.jp
    }
    \IEEEauthorblockA{
        \IEEEauthorrefmark{2}D3 Center, %
        The University of Osaka, Osaka, Japan, %
        Email: shimonishi.cmc@osaka-u.ac.jp
    }
}

\maketitle

\begin{abstract}
Smart home automation that adapts to a user's emotional state can enhance psychological safety in daily living environments.
This study proposes an emotion-aware automation framework guided by the emotional Biologically Inspired Cognitive Architecture (eBICA), which integrates appraisal, somatic responses, and behavior selection.
We conducted a proof-of-concept experiment in a pseudo-smart-home environment, where participants were exposed to an anxiety-inducing event followed by a comfort-inducing automation.
State anxiety (STAI-S) was measured throughout the task sequence. The results showed a significant reduction in STAI-S immediately after introducing the avoidance automation, demonstrating that emotion-based control can effectively promote psychological safety. Furthermore, an analysis of individual characteristics suggested that personality and anxiety-related traits modulate the degree of relief, indicating the potential for personalized emotion-adaptive automation. Overall, this study provides empirical evidence that eBICA-based emotional control can function effectively in smart home environments and offers a foundation for next-generation affective home automation systems.
\end{abstract}


\section{Introduction}
With the widespread adoption of IoT devices, smart-home automation systems that integrate and control home appliances are increasingly being utilized.
Conventional automation predominantly relies on rule-based control, schedules tied to time or location, or predictive adjustments based on users' behavioral patterns~\cite{Yang18SmartHomeAutomationBasedOnBehavior,Benhaddou2019AutonomationByBehavior}, and these approaches have mainly been developed to improve comfort and efficiency.
However, in home environments where people live, maintaining not only physical comfort and safety but also psychological security is essential.
In particular, home automation that directly addresses emotional-state transitions—such as reducing anxiety and fostering a sense of reassurance—has not been sufficiently explored. There is a need for a framework capable of selecting appropriate control actions based on the user's emotional state and evaluating how those actions subsequently alter the user's emotions.

Most existing studies visualize estimated emotions or trigger simple
device reactions, and only a few have examined whether automation
can actually alter users' emotional states~\cite{Kalam20EmotionPredictionBioFeedback,Ghadekar24HomeEmotionRecognition,Kaushik22SmartHomeSurveillanceAlert}
Although some
works modify lighting based on facial or vocal emotion recognition~\cite{Patil21HomeAutomationLED,Boon15Eye2H,Jaihar2020EmotionAutomation}, the psychological impact of such interventions has rarely
been validated.

\begin{figure}[t]
    \centering
    \includegraphics[width=.8\linewidth]{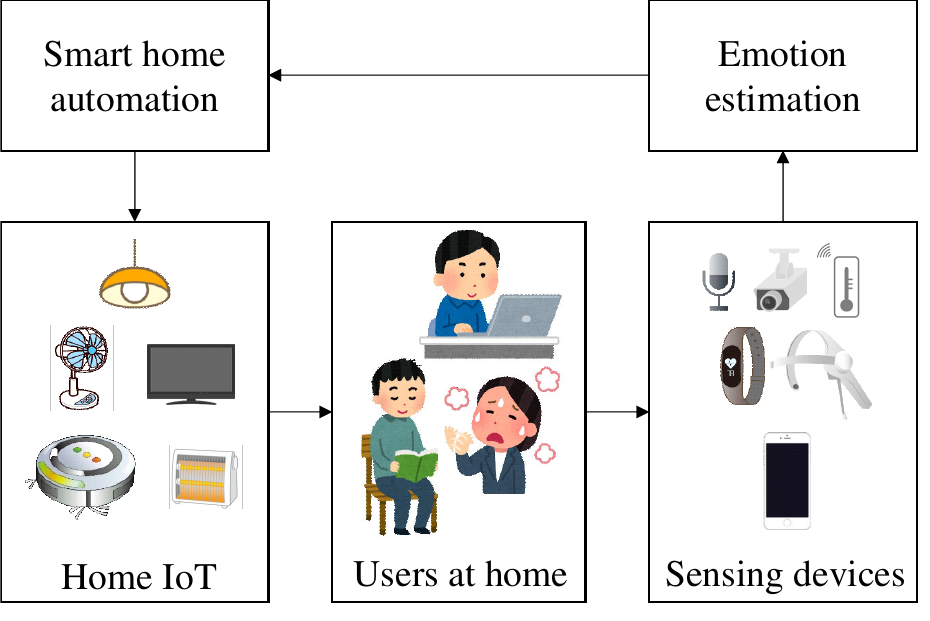}
    \caption{Overview of the proposed framework.}
    \label{fig:OverviewFramework}
\end{figure}

In this study, we propose a home-automation framework that estimates a user's emotional state from observations obtained in a smart-home environment and dynamically controls appliance behaviors according to that state (Fig.~\ref{fig:OverviewFramework}).
In particular, we adopt the emotional Biologically Inspired Cognitive Architecture (eBICA) model~\cite{Samsonovich20eBICA} as a design guideline, which integrates emotional appraisal and action selection. Based on the estimated emotion, the framework selects the most appropriate action (automation) to execute.
The eBICA model is well suited to this task, as it provides a mathematical representation of the processes underlying emotional appraisal, physiological responses, and behavior selection, enabling the dynamic handling of emotion transitions.

In this paper, we focus on reassurance and anxiety as representative emotions related to psychological safety in home environments, and examine whether the proposed framework can reduce anxiety and enhance reassurance.
Specifically, we construct an anxiety-inducing scenario based on a plausible household hazard and evaluate how emotional states dynamically change when a reassurance-providing automation is presented as an intervention.
Furthermore, by analyzing individual factors that influence changes in reassurance, we identify user characteristics that are more responsive to such interventions and discuss the potential for extending the eBICA-based framework toward personalized emotional adaptation.

The contributions of this study are as follows:
\begin{itemize}
    \item We propose an emotion-adaptive smart-home automation framework guided by the eBICA model.
    \item We conduct a proof-of-concept (PoC) demonstration of an automation designed to reduce anxiety and enhance reassurance.
    \item We analyze individual differences in emotional responses based on personality traits and discuss directions for user modeling grounded in these findings.
\end{itemize}

The remainder of this paper is organized as follows.
Section~\ref{sec:proposal} presents an overview of the proposed framework.
Section~\ref{sec:experiment} describes the design of the PoC experiment, and Section~\ref{sec:evaluation} reports the evaluation results and analyses of individual differences.
Section~\ref{sec:discussion} provides the discussion, and Section~\ref{sec:conclusion} concludes the paper.

\section{Proposed emotion-aware smart home automation framework based on eBICA} \label{sec:proposal}
\begin{figure}[t]
    \centering
    \includegraphics[width=.98\linewidth]{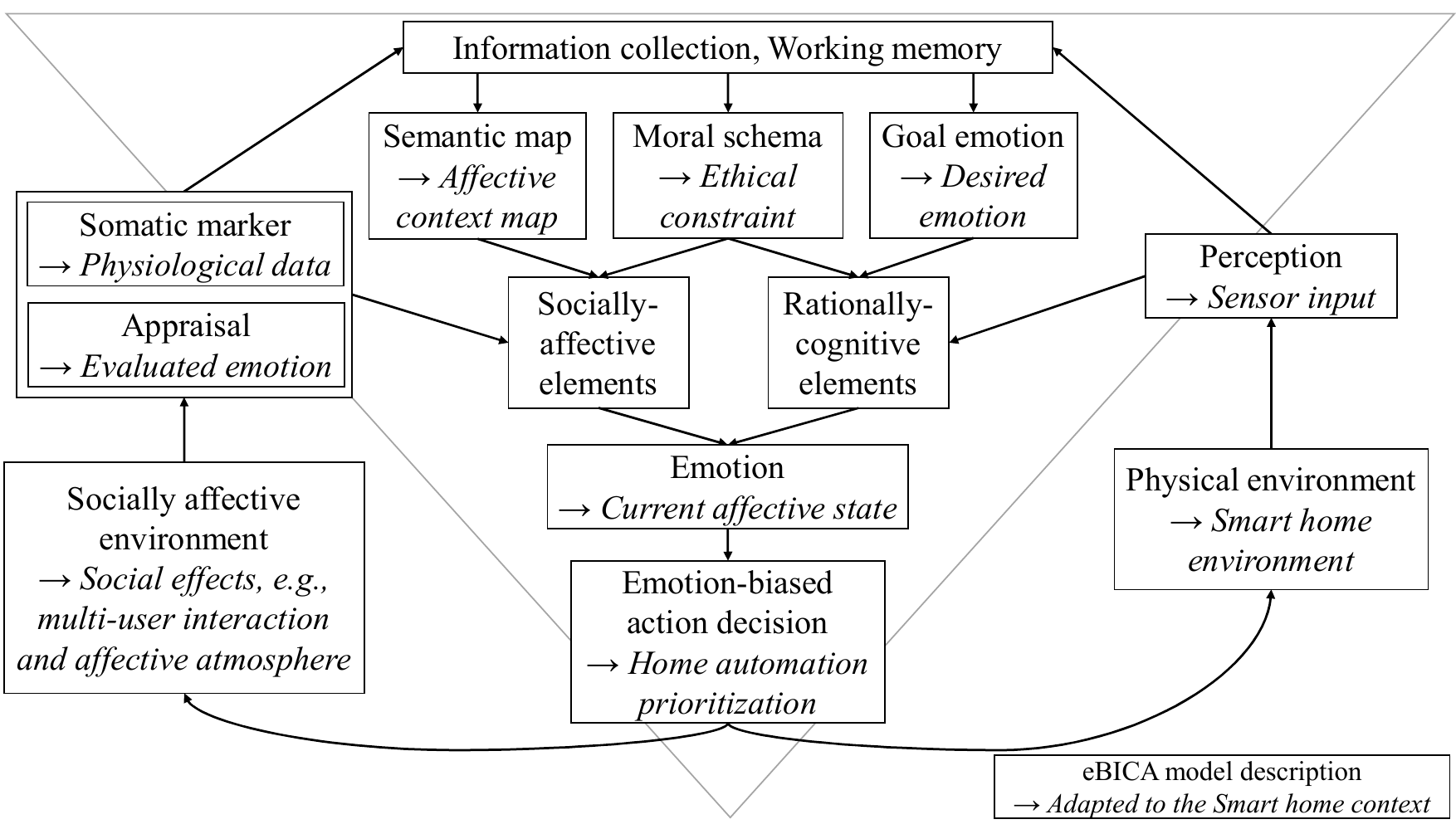}
    \caption{Overview of the eBICA model, redrawn by authors based on~\cite{Samsonovich20eBICA}.}
    \label{fig:OverviewOfeBICAFramework}
\end{figure}

The proposed automation framework is guided by the emotional Biologically Inspired Cognitive Architecture (eBICA)~\cite{Samsonovich20eBICA}, which integrates appraisal, somatic markers, emotional updating, and behavior selection within a unified structure.
While prior applications of eBICA have primarily focused on virtual agents and socially affective systems~\cite{Veselov19eBICAbasedNPC,Bogatyreva18eBICAbasedVirtualPet}, its applicability to real smart home environments has not been explored.
This gap motivates our investigation of whether eBICA's affective mechanisms can operate effectively in a physical home-like setting.

eBICA provides an affective–behavior loop in which appraisal, somatic markers, emotional state, and action selection interact. In this study, we adopt only the components necessary for emotion-driven control: S, A, F, E, and B, as shown in Fig.~\ref{fig:OverviewOfeBICAFramework}.
Higher-level elements such as semantic mapping or moral schema are excluded because they are not required for this PoC. Table~\ref{tab:eBICA_variables} summarizes only the variables used here.
Each variable is briefly defined with its role in the simplified PoC setting.

\begin{table}[t!]
    \centering
    \caption{Symbols and their meanings in the eBICA model\label{tab:eBICA_variables}}
    \begin{tabularx}{\linewidth}{c p{0.21\linewidth} X}
        \toprule
        \textbf{Symbol} & \textbf{Original meaning in eBICA} & \textbf{Role in this study} \\
        \midrule
        $A$ & Appraisal 
            & An evaluation vector (e.g., valence, arousal) used to estimate the user's comfort and reassurance levels. \\
        $S$ & Somatic Marker
            & Physiological indicators such as facial expressions, heart rate, and bodily reactions reflecting responses to the environment. \\
        $F$ & Goal Emotion
            & The desired emotional state of the user (e.g., to feel reassured or calm). \\
        $E$ & Emotion
            & The estimated emotional value at each moment, updated by the mathematical model (e.g., comfort or reassurance level). \\
        $B$ & Emotion-biased action decision
            & A metric that determines which automation should be prioritized to move the emotional state toward the goal emotion. \\
        $a$, $a^*$ & Action (Home automation)
            & From candidate automation $a$, the optimal action $a^*$ is selected and executed based on $B(a)$. \\
        \bottomrule
    \end{tabularx}
\end{table}

In the eBICA model, the internal emotional state $E$ updates according to the following rule:
\begin{equation}
    E(t+1) = (1-r)E(t) + r \bigl( c_1 A(t) + c_2 S(t) + c_3 F(t) \bigr),
\end{equation}
where $A(t)$, $S(t)$, and $F(t)$ denotes the appraisal, somatic markers, and the goal emotion, respectively, and $r$ and $c_i$ are tuning parameters.
This formulation expresses how the current emotional state integrates cognitive evaluations, physiological cues, and the desired emotional target.

To determine which automation should be executed, the bias value for each candidate action (automation) $a$ is computed as:
\begin{equation}\label{eq:Ba}
    B(a) = - \| F - A(a) \|,
\end{equation}
and the action that maximizes this bias is selected:
\begin{equation}
    a^* = \arg\max_a B(a).
\end{equation}
This mechanism prioritizes automations predicted to shift the user's emotional state toward the desired goal.

Through this affective–behavior loop, the system dynamically updates the user's emotional state based on the executed automation, enabling adaptive control grounded in eBICA. In the subsequent PoC experiment, we examine whether introducing an appropriate automation can immediately modulate the emotional state in accordance with this loop structure.

\section{Experiment} \label{sec:experiment}
In a pseudo–smart-home environment, anxiety is induced by causing a shelf to fall, after which an avoidance system is presented to evaluate changes in reassurance.
A total of 40 participants are assigned to a main experiment group, a control group, and a long-duration auxiliary group, and the transitions of state anxiety are compared across these groups.

\subsection{Ethical and Safety Considerations}
This experiment was conducted with the approval of the Research Ethics Committee of the Graduate School of Information Science and Technology, The University of Osaka (Approval No.: 202403).
Participants were informed that they could discontinue the experiment at any time, and a debriefing was provided afterward to explain that both the shelf-collapse event and the avoidance system were preplanned components of the study.
For safety, the direction and position of the falling shelf were carefully adjusted in advance to ensure that it would not come into contact with participants.

\subsection{Participants}
A total of 40 participants (ages 18--29; mean = 21.0, SD = 2.5; 17 males and 23 females) were recruited through a university bulletin board.
Participants were randomly assigned to one of three groups: 15 to the main experiment group in which the avoidance automation was introduced, 10 to the control group without the automation, and 15 to the long-duration auxiliary group.
Eligibility criteria required that participants be native Japanese speakers residing in Japan and have no visual or auditory impairments.  
To avoid risks related to anxiety or allergic reactions, individuals with metal or alcohol allergies, cardiovascular conditions, or those taking headache medication, sleeping pills, antidepressants, or antipsychotic drugs were excluded.
Participants who completed the experiment received a reward equivalent to 2,500 JPY.

\subsection{Experimental protocol}
The experiment consisted of an anxiety-inducing event and a reassurance-providing condition with cognitive tasks (T) and questionnaires (Q) as shown in Table~\ref{tab:ExpTimeline_S}.
Anxiety is induced by causing a shelf to fall during task T3, Q4–Q6 are recorded as the anxiety condition, and reassurance is provided by introducing an avoidance automation after Q6.
Subsequently, the avoidance system and an introductory video explaining is presented, after which the robot operated with avoidance behavior enabled; Q7–Q10 are recorded as the reassurance condition.

In the control group, a household appliance advertisement was shown instead of the introductory video, while all other procedures were identical.

The long-duration auxiliary experiment followed the same procedure as the main experiment, with an extended overall duration.
Specifically, the response time for each Q was set to 2 minutes, and each cognitive task was extended to 4 minutes.
In this condition, the robot was not operated during T1; the shelf was made to fall 2 minutes after the start of T4; and the introduction and activation of the avoidance automation were conducted between Q7 and Q8.

\begin{table}[htb]
    \centering
    \caption{Experimental procedure\label{tab:ExpTimeline_S}}
    \begin{threeparttable}
    \begin{tabularx}{\linewidth}{p{0.15\linewidth} p{0.32\linewidth} p{0.27\linewidth} p{0.1\linewidth}}
        \toprule
        \textbf{Section} & \textbf{Activity details} & \textbf{Robot vacuum} & \textbf{\makecell[l]{Time\\(min)}} \\
        \midrule
        Preparation
        & \makecell[l]{Informed consent, pre-\\questionnaire, task practice} & & 30 \\
        \midrule
        \makecell[l]{Neutral\\condition}
        & Q1 & & 1.5 \\
        & T1 & Working\textsuperscript{*} & 3 \\
        & Q2 & & 1.5 \\
        & T2 & Working\textsuperscript{*} & 3 \\
        & Q3 & & 1.5 \\
        \makecell[l]{Inducing\\discomfort}
        & T3 & \makecell[l]{Working\textsuperscript{*} and push-\\ing the shelf down\textsuperscript{**}} & 3 \\
        \makecell[l]{Discomfort\\condition}
        & Q4 & & 1.5 \\
        & T4 & Working\textsuperscript{*} & 3 \\
        & Q5 & & 1.5 \\
        & T5 & Working\textsuperscript{*} & 3 \\
        & Q6 & & 1.5 \\
        \makecell[l]{Inducing\\comfort}& \makecell[l]{Watching\\introduction video of\\the avoidance system} & & 5 \\
        Comfort
        & Q7 & & 1.5 \\
        & T6 & Working\textsuperscript{*} & 3 \\
        & Q8 & & 1.5 \\
        & T7 & Working\textsuperscript{*} & 3 \\
        & Q9 & & 1.5 \\
        & T8 & Working\textsuperscript{*} & 3 \\
        & Q10 & & 1.5 \\
        \midrule
        \makecell[l]{Post-\\experiment}
        & \makecell[l]{Revealing,\\post-questionnaire} & & 30 \\
        \bottomrule
    \end{tabularx}
    \begin{tablenotes}
        \footnotesize
        \item[] Q: STAI-S questionnaire; T: cognitive task (receipt categorization~\cite{Ishii18QuantitativeCognitiveTask}).
        \item[\textsuperscript{*}] The vacuum robot runs from 30~s to 150~s within the 3~min task.
        \item[\textsuperscript{**}] The vacuum robot pushes the shelf off 90~s after task initiation.
    \end{tablenotes}
    \end{threeparttable}
\end{table}

\begin{figure}[t]
    \centering
    \includegraphics[width=.98\linewidth]{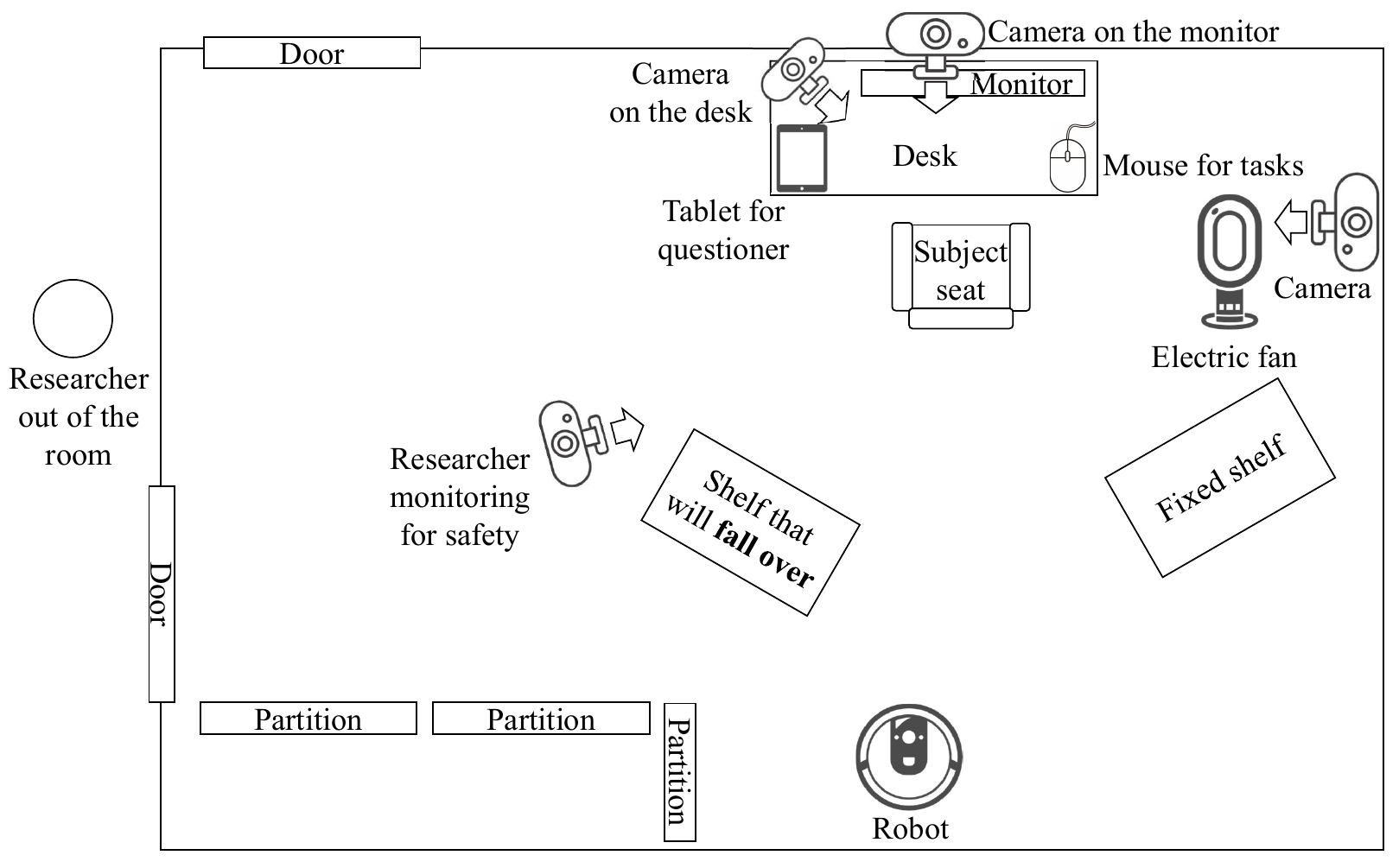}
    \caption{Layout of the experimental environment}
    \label{fig:ExperimentalEnvironment}
\end{figure}

The experimental smart home environment shown in Fig.~\ref{fig:ExperimentalEnvironment} was constructed in a classroom.
A robot vacuum, shelves, home appliances, and a desk and chair were arranged in the space, and task instructions and videos were presented on a monitor placed in front of the participant.

The robot vacuum followed a predefined path within the room and was operated so that it made contact with shelves and furniture under the normal condition.
After the avoidance automation was introduced, the robot followed the same path but was controlled to avoid contact with the shelves.

\subsection{Acquiring data}

Data acquisition was conducted using three cameras placed in the environment to record participants' facial expressions and behaviors during the experiment.
Physiological data, including heart rate, are measured using an EEG device and a wristband-type vital sensor.

For psychological measures, the Japanese version of the State-Trait Anxiety Inventory, State form (STAI-S)~\cite{Spielberger70AnxietyIndexSTAIBase,Iwamoto89AnxietyIndexSTAIJP}, a standard scale for assessing state anxiety, is administered at each measurement point.
In addition, demographic information (gender and age), personality traits assessed using the TIPI-J~\cite{GOSLING03TIPI,Oshio12TIPI-J}, and trait anxiety scores from the STAI-T~\cite{Iwamoto89AnxietyIndexSTAIJP} were collected through a pre-experiment questionnaire and used for analyzing individual differences in reassurance responses.

\subsection{Avoidance Automation Provided}
In the avoidance-automation condition, a video was presented explaining that a collision-avoidance system linked to a monitoring camera had been introduced, and during the subsequent tasks the robot vacuum operated so as to avoid shelves and walls.
The actual control was performed manually by the experimenter, who adjusted the robot's movement to follow the same path as in the normal condition while avoiding contact with the shelves.
In the control condition, a household appliance advertisement was shown at the corresponding timing, and the robot operated along its usual path with minor contacts.  
Note that the avoidance behavior was implemented conceptually and manually controlled by the experimenter.

\section{Evaluation} \label{sec:evaluation}
As a PoC demonstration of the proposed framework, we evaluated the effectiveness of the reassurance-providing automation.
It should be noted that the estimation processes of appraisal and somatic markers in the eBICA model (i.e., the computation of $A$ and $S$), as well as the optimization of the automation timing $t$, are outside the scope of this study.  
Accordingly, the present evaluation focuses on empirically examining how the automation influences emotional changes.
This PoC focuses on the psychological effect of reassurance-oriented intervention rather than on full automation; the participants were not disclosed the robot control mode (manual or automatic).

\subsection{Application to the eBICA model}
In this evaluation, the components of the eBICA model are simplified to match the PoC experimental setting.
The appraisal value $A(t)$ is treated as a one-dimensional emotional evaluation derived from the STAI-S score at each measurement point, and no estimation is performed from facial or physiological signals.  
We exclude physiological data corresponding to $S(t)$ from the present analysis.
The goal emotion $F(t)$ is set as a constant representing a non-anxious state.
The action $a$ represents a binary choice indicating whether the avoidance automation is introduced.  
For the PoC implementation, the update equation is simplified to $E(t+1)=A(t)$, and the evaluation examines whether anxiety decreases immediately after the automation is introduced.

\subsection{Evaluation of reassurance changes by the automation}

\subsubsection{Method}
Using the STAI-S scores at each questionnaire point, we calculated the change in anxiety ($\Delta$STAI) before and after the anxiety-inducing event, as well as the introduction of the avoidance automation.
For both the main experiment group and the control group, we computed the mean, median, and standard error of these changes, and compared statistical differences within and between the groups.

\subsubsection{Results} \label{sec:evaluation_result_of_stais}

\begin{figure}[t]
    \centering
    \includegraphics[width=0.8\linewidth,clip]{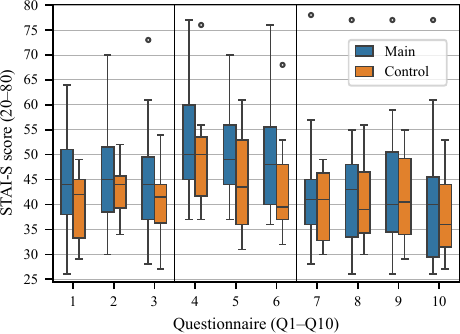}
    \caption{Transition of STAI-S scores in the main ($n=15$) and control ($n=10$) experiments. A shelf-collapse event (anxiety induction) occurred between Q3 and Q4. Between Q6 and Q7, the avoidance system (reassurance-providing automation) was introduced in the main group, whereas an unrelated advertisement video was presented in the control group.}
    \label{fig:Q1-10_C_S}
\end{figure}
\begin{figure}[t]
    \centering
    \includegraphics[width=0.8\linewidth,clip]{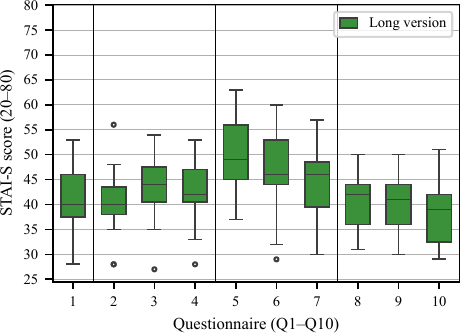}
    \caption{Transition of STAI-S scores in the long-duration experiment ($n=15$). A shelf-collapse event occurred between Q4 and Q5, followed by the introduction of the avoidance system between Q7 and Q8.}
    \label{fig:Q1-10_R}
\end{figure}
\begin{table}[t]
    \centering
    \caption{Changes in STAI-S scores during discomfort induction (shelf collision) and the automation to induct comfort (avoidance)}
    \label{tab:STAI_delta_summary}
    \begin{tabularx}{\columnwidth}{Xlccccc}
    \toprule
    Event & Group & n & Mean$\Delta\pm$SD & Med.$\Delta$ & $p$ & $r$ \\
    \midrule
    \multirow{3}{*}{\shortstack[l]{Shelf\\collision}}
    & Main & 15 & 7.9 $\pm$ 5.8 & 9.0 & $<$.001 & .85 \\
    & Long & 15 & 7.3 $\pm$ 5.4 & 7.0 & $<$.001 & .88 \\
    & Control & 10 & 9.7 $\pm$ 10.8 & 6.5 & .004 & .89 \\
    \midrule
    \multirow{2}{*}{\shortstack[l]{Avoidance}}
    & Main & 15 & -6.3 $\pm$ 9.6 & -3.0 & .010 & .69 \\
    & Long & 15 & -3.7 $\pm$ 3.6 & -4.0 & .002 & .75 \\
    \multirow{1}{*}{\shortstack[l]{CM}}
    & Control &10 & -3.3 $\pm$ 7.8 & -1.5 & .195 & .50 \\
    \bottomrule
    \multicolumn{7}{l}{\footnotesize Med.: Median; $\Delta$: score difference (after–before).}
    \end{tabularx}
\end{table}

Fig.~\ref{fig:Q1-10_C_S} shows the transition of STAI-S scores in the main and control groups.
From Q3 to Q4, the shelf-collapse event caused a sharp increase in anxiety, followed by a gradual decrease over time.
However, in the main group, a clear and immediate reduction in anxiety was observed immediately after the introduction of the avoidance automation between Q6 and Q7.
This pattern suggests that the avoidance automation substantially accelerated the recovery from anxiety.

A similar trend was observed in the Long group (Fig.~\ref{fig:Q1-10_R}).  
STAI-S increased markedly immediately after the anxiety induction and then gradually decreased over time.  
Following the introduction of the avoidance automation, a distinct additional decrease was observed, after which the anxiety levels remained relatively stable.  
Moreover, the decrease in the Main group was larger than that in the Long group, indicating that providing the reassurance automation earlier may shorten the period during which anxiety remains elevated.

Table~\ref{tab:STAI_delta_summary} summarizes the differences in STAI-S scores before and after each event.
For the anxiety-inducing event, Shelf collision, all groups showed a significant increase in STAI-S (Median $\Delta$ = 6.5--9.0, $p<.005$, $r=.85$--$.89$), confirming that the event served as a strong anxiety inducer.

For the reassurance-providing event, Avoidance, both the Main and Long groups showed significant reductions in STAI-S (Median $\Delta=-3.0$ and $-4.0$, $p = .010$ and $.002$, $r = .69$ and $.75$).  
In contrast, the Control group, which viewed an unrelated commercial, did not show a significant decrease (Median $\Delta = -1.5$, $p = .195$, $r = .50$).  
This strongly indicates that the reductions observed in the Main and Long groups were not solely due to natural recovery over time, but were instead driven by the active environmental intervention provided by the avoidance automation.

Overall, across all groups, the general time-dependent pattern of gradual natural recovery after anxiety induction and the additional, pronounced decrease immediately following the introduction of the avoidance automation were consistently observed.  
These results empirically demonstrate that emotion-based automation can accelerate psychological recovery and enhance reassurance.  
This observation aligns with the eBICA framework, in which behavior selection ($B$) contributes to the regulation of emotional state ($E$), thereby providing empirical support for emotion-driven smart-home automation.

On the other hand, as shown in Table~\ref{tab:STAI_delta_summary}, the standard deviations of $\Delta$STAI were relatively large for both the anxiety-inducing and reassurance-providing events, indicating substantial variability in individual responses.  
This suggests that susceptibility to anxiety and the effectiveness of reassurance automation may vary across individuals.  
The next section analyzes the relationship between these individual characteristics and emotional responses.

\subsection{Individual differences in responses to the automation}
While the avoidance automation contributed to reducing anxiety overall, the large standard deviations of the change values indicate clear individual differences in its effectiveness.  
This suggests that even when presented with the same automation, some users may experience reassurance more readily, whereas others may show only limited improvement.

Accordingly, when applying eBICA-based emotion-adaptive automation in real environments, it is necessary to consider that effectiveness may vary across individuals.
In this section, we examine factors contributing to these differences, focusing on gender, age, trait anxiety (STAI-T), and personality traits (the Big Five assessed via TIPI-J).

Previous studies have reported that younger individuals in Japan tend to experience higher levels of anxiety~\cite{Iwamoto89AnxietyIndexSTAIJP}, and that greater agreeableness and extraversion are associated with higher technology acceptance~\cite{Ozbek14PersonalityTechnologyAcceptance}.  
These findings suggest that personality-related factors may influence how users perceive the reassurance provided by the avoidance automation.

Based on these considerations, we analyze which personal characteristics contribute to more effective responses to the automation, and discuss design guidelines for personalized emotion-adaptive automation grounded in the eBICA model.

\subsubsection{Method}
To analyze individual differences in psychological responses to the automation, we calculated the change in STAI-S scores for both the anxiety-inducing event and the reassurance-providing automation.  
Specifically, we defined the increase in anxiety before and after the shelf-collapse event as $\Delta$Anx (Anxiety), and the decrease in anxiety before and after the introduction of the avoidance automation as $\Delta$Rel (Relief).

To examine how these change values relate to participants' individual characteristics (age, trait anxiety as measured by STAI-T, and Big Five personality traits assessed using TIPI-J), Pearson correlation coefficients were computed separately for males and females.  
The TIPI-J is a Japanese personality inventory that assesses five domains—extraversion, agreeableness, conscientiousness, neuroticism, and openness—using 10 items on a 7-point scale~\cite{Oshio12TIPI-J}.  
In this study, correlations are evaluated for each trait dimension.

One participant whose STAI-S responses were extremely biased (responding almost exclusively with 1 or 2 across the 20 items) was excluded from the analysis, as such patterns substantially undermine the reliability of the scale.

\subsubsection{Results}

Table~\ref{tab:corr_delta1_delta2_by_gender} shows the correlations between
participant characteristics and emotional change values for both the
anxiety-inducing event ($\Delta$Anx) and the reassurance-providing
automation ($\Delta$Rel), separated by gender.

\newcolumntype{R}{>{\raggedleft\arraybackslash}p{0.07\linewidth}}
\newcolumntype{V}{>{\raggedleft\arraybackslash}p{0.04\linewidth}}
\begin{table}[t]
    \centering
    \caption{Correlations between participant characteristics and emotional responses to anxiety induction $\Delta$Anx and relief automation $\Delta$Rel separated by gender.}
    \label{tab:corr_delta1_delta2_by_gender}
    \begin{tabular}{clVRRVRR}
    \toprule
    \multirow{2}{*}{\shortstack[c]{Gender}} & \multirow{2}{*}{\shortstack[c]{Predictor}} & \multicolumn{3}{c}{$\Delta$Anx} & \multicolumn{3}{c}{$\Delta$Rel} \\
     & & \multicolumn{1}{c}{n} & \multicolumn{1}{c}{r} & \multicolumn{1}{c}{p} & \multicolumn{1}{c}{n} & \multicolumn{1}{c}{r} & \multicolumn{1}{c}{p} \\
    \midrule
    M & Age       & 17 & \textbf{-.43} & \textbf{.085} & 15 & -.02 & .933 \\
    M & STAI-T    & 17 &  .12 & .643 & 15 & -.01 & .962 \\
    M & TIPI-J Q1 & 17 & -.20 & .437 & 15 & -.03 & .905 \\
    M & TIPI-J Q2 & 17 & -.17 & .511 & 15 & -.00 & 1.00 \\
    M & TIPI-J Q3 & 17 &  .21 & .423 & 15 & -.18 & .514 \\
    M & TIPI-J Q4 & 17 &  .01 & .975 & 15 & -.31 & .259 \\
    M & TIPI-J Q5 & 17 & -.10 & .716 & 15 &  .36 & .186 \\
    M & TIPI-J Q6 & 17 &  .33 & .197 & 15 & -.26 & .351 \\
    M & TIPI-J Q7 & 17 &  .20 & .435 & 15 & -.29 & .288 \\
    M & TIPI-J Q8 & 17 & -.33 & .197 & 15 &  .19 & .503 \\
    M & TIPI-J Q9 & 17 & \textbf{.46} & \textbf{.064} & 15 & -.00 & 1.00 \\
    M & TIPI-J Q10& 17 &  .12 & .661 & 15 & -.19 & .509 \\
    \midrule
    F & Age       & 23 &  .14 & .519 & 15 & \textbf{-.49} & \textbf{.064} \\
    F & STAI-T    & 23 & \textbf{-.36} & \textbf{.091} & 15 &  .35 & .208 \\
    F & TIPI-J Q1 & 23 & -.08 & .721 & 15 & -.08 & .778 \\
    F & TIPI-J Q2 & 23 & -.26 & .233 & 15 &  .25 & .379 \\
    F & TIPI-J Q3 & 23 & -.07 & .736 & 15 &  .03 & .912 \\
    F & TIPI-J Q4 & 23 & -.07 & .756 & 15 &  .38 & .163 \\
    F & TIPI-J Q5 & 23 &  .08 & .717 & 15 &  .30 & .278 \\
    F & TIPI-J Q6 & 23 & -.20 & .358 & 15 &  .30 & .280 \\
    F & TIPI-J Q7 & 23 &  .16 & .471 & 15 & -.10 & .726 \\
    F & TIPI-J Q8 & 23 & -.01 & .969 & 15 & -.00 & .988 \\
    F & TIPI-J Q9 & 23 &  .34 & .111 & 15 & \textbf{-.59} & \textbf{.021} \\
    F & TIPI-J Q10& 23 &  .05 & .814 & 15 & -.19 & .496 \\
    \bottomrule
    \end{tabular}
\end{table}

Regarding the anxiety-inducing event $\Delta$Anx, both shared tendencies
and gender-specific patterns were observed.
For males, age showed a moderate correlation with $\Delta$Anx
($r=-.43$, $p=.085$), indicating that younger males tended to show
larger anxiety increases.
Emotional stability, reflected in TIPI-J Q9, showed a positive correlation
with $\Delta$Anx ($r=.46$, $p=.064$), suggesting that males who rated
themselves as more emotionally stable showed larger anxiety increases.

In contrast, among females, no correlations were found with TIPI-J
personality items.
However, trait anxiety (STAI-T) showed a negative correlation with
$\Delta$Anx ($r = -.36$, $p = .091$), indicating that females with higher
baseline anxiety showed smaller anxiety increases
during the shelf-collapse.

Next, for the reassurance-providing automation $\Delta$Rel, distinct
gender differences emerged.  
For males, none of the predictors showed meaningful correlations with $\Delta$Rel ($|r|<.36$, $p>.18$), indicating that the effectiveness of the avoidance automation was relatively uniform across male participants.

For females, however, emotional stability (TIPI-J Q9) showed a significant
correlation with $\Delta$Rel ($r = -.59$, $p = .021$), and age also showed a
moderate association ($r = -.49$, $p = .064$).  
Because $\Delta$Rel represents the reduction in anxiety, these results
suggest that older females and those with higher emotional stability
experienced stronger reassurance effects from the automation.

\section{Discussion} \label{sec:discussion}
This study is the first attempt to examine the cyclical structure of
emotion updating and action selection in the eBICA model within a real
smart-home environment. A consistent pattern was observed in which the
shelf-collapse event caused a rise in anxiety, followed by a clear
decrease immediately after the introduction of the avoidance automation.
These results demonstrate that the appraisal–emotion–action loop assumed
in eBICA can be applied to psychological state transitions occurring in
actual environments.

Because anxiety naturally decreases over time and the explanatory
video and avoidance behavior were presented together, the individual
effects cannot be isolated; future studies should clarify these factors.

The analysis of individual differences revealed that age and emotional stability influenced responses to both the anxiety-inducing event and the avoidance automation. In particular, emotional stability was strongly associated with reassurance recovery in females, whereas the effect tended to be more uniform among males.
Such individual characteristics could be used as an initial and lightweight user profile for future personalization of eBICA-based emotion-adaptive control.
However, the subgroup sample sizes for the personality analysis were limited, and the observed relationships should be interpreted as indicative trends.

In this study, emotional states were evaluated solely using the
self-report STAI-S, and continuous or real-time estimation was not
performed. Future work should integrate physiological indicators such as
heart rate and facial expressions, and optimize the eBICA update rules
and weights in a data-driven manner to enable real-time
feedback-based adaptation. Incorporating elements related to
social context and semantic reasoning represents an important avenue
for future extensions.

\section{Conclusion} \label{sec:conclusion}
In this study, we conducted a proof-of-concept demonstration of an
emotion-adaptive automation based on the eBICA model and showed that the
avoidance control significantly reduced state anxiety.  
We also confirmed that changes in reassurance were influenced by factors
such as age and emotional stability, highlighting the importance of
personalized adaptive control.
Future work will pursue real-time emotional estimation and adaptive control.

\bibliographystyle{IEEEtran}

\end{document}